\documentclass[oribibl]{llncs}
\usepackage{llncsdoc}

\usepackage[T1]{fontenc}
\usepackage[american]{babel}
\usepackage{hyperref}
\usepackage{cite}
\usepackage{fancybox}
\usepackage{amsmath}
\usepackage{amsfonts}
\usepackage{amssymb}
\usepackage{graphicx}
\usepackage{framed}
\usepackage{tabularx}

\usepackage{amssymb}
\usepackage{textcomp}
\usepackage{longtable}
\usepackage{multirow}
\usepackage{pifont}
\usepackage{changepage}
\usepackage{listings}
\usepackage{url}
\usepackage{caption}
\usepackage{xspace}
\usepackage{xtab}
\usepackage[utf8]{inputenc}
\usepackage[T1]{fontenc}
\usepackage{ltablex,booktabs}
\usepackage{cite}
\usepackage[toc,page]{appendix}
\usepackage{float}
\usepackage{hyperref}
\hypersetup{
	colorlinks = true,
	linkcolor = black
}
\hypersetup{
	citecolor=black
}
\hypersetup{
	urlcolor=black
}
\DeclareGraphicsExtensions{.pdf}

\addto\captionsenglish{
\renewcommand{\figurename}{Fig.}
}

\title{Checklists to Support Test Charter Design in Exploratory Testing}

\author{Ahmad Nauman Ghazi \and Ratna Pranathi Garigapati \and Kai Petersen}

\institute{Blekinge Institute of Technology, Karlskrona, Sweden \\
\email{nauman.ghazi@bth.se, pranathi.r8@gmail.com, kai.petersen@bth.se}
}

\date{}

\begin{document}

\maketitle

\begin{abstract}
During exploratory testing sessions the tester simultaneously learns, designs and executes tests. The activity is iterative and utilizes the skills of the tester and provides flexibility and creativity.Test charters are used as a vehicle to support the testers during the testing. The aim of this study is to support practitioners in the design of test charters through checklists. We aimed to identify factors allowing practitioners to critically reflect on their designs and contents of test charters to support practitioners in making informed decisions of what to include in test charters. The factors and contents have been elicited through interviews. Overall, 30 factors and 35 content elements have been elicited.

\end{abstract}

\section{Introduction}\label{sec:intro}

James Bach defines exploratory testing as simultaneous learning, test design and test execution~\cite{bach2003exploratory}. Existing literature reflects that ET is widely used for testing complex systems as well and is perceived to be flexible in all types of test levels, activities and phases \cite{ghazi2015heterogeneous}\cite{pfahl2014exploratory}. In the context of quality, ET has amassed a good amount of evidence on overall defect detection effectiveness, cost effectiveness and high performance for detecting critical defects~\cite{afzal2015experiment}~\cite{itkonen2011empirical}~\cite{itkonen2014test}~\cite{itkonen2005exploratory}~\cite{pfahl2014exploratory}.  Session-based test management (SBTM) is an enhancement to ET. SBTM incorporates planning, structuring, guiding and tracking the test effort with good tool support when conducting ET \cite{bach2007rapid}.

A test charter is a clear mission for the test session and a high level plan that determines what should be tested, how it should be tested and the associated limitations. A tester interacts with the product to accomplish a test mission or charter and further reports the results \cite{bach2003exploratory}.  The charter does not pre-specify the detailed test cases which are executed in each session. But, a total set of charters for an entire project generally include everything that is reasonably testable. The metrics gathered during the session are used to track down the testing process more closely and to make instant reports to management \cite{itkonen2005exploratory}. Specific charters demand more effort in their design whilst providing better focus. A test session often begins with a charter which forms the first part of the scannable session sheet or the reviewable result. Normally, a test charter includes the mission statement and the areas to be tested in its design. 

 Overall, the empirical evidence of how test charters are designed and how to achieve high quality test charters are designed are scarce. High quality test charters are useful, accurate, efficient, adaptable, clear, usable, compliant, and feasible  \cite{bach2007rapid}. In this study we make a first step towards understanding test charter design by exploring the factors influencing the design choices, and the elements that could be included in a test charter. This provides the foundation for further studies investigating which elements actually lead to the quality criteria described by Bach \cite{bach2007rapid}. We make the following contributions:

\begin{itemize}
\item[C1:] Identify and categorize the influential factors that practitioners consider when designing test charters. 
\item[C2:] Identify and categorize the possible elements of a test charter.
\end{itemize}

The remainder of the paper is structured as follows: Section~\ref{sec:related} presents the related work. Section~\ref{sec:rm} outlines the research method, followed by the results in Section~\ref{sec:results}. Finally, in Section~\ref{sec:conclusion}, we present the conclusions of this study.

\section{Related work}\label{sec:related}
 
 \textit{Test charters}, which are an SBTM element plays a major role in guiding inexperienced testers. The charter is a test plan which is usually generated from a test strategy. The charters include ideas that guide the testers as they test. These ideas are partially documented and are subject to change as the project evolves \cite{bach2007rapid}. SBTM echoes the actions of testers who are well experienced in testing and charters play a key role in guiding the inexperienced testers by providing them with details regarding the aspects and actions involved in the particular test session \cite{bach2000session}. 
 
The context of the test session plays a great role in determining the design of test plan or the charter \cite{bach2007rapid}. Key steps to achieve context awareness are, for example, understanding the project members and the way they are affected by the charter, and understanding work constraints and resources. When designing charters Bach \cite{bach2007rapid} formulated specific goals, in particular finding significant tests quicker, improving quality, and increasing testing efficiency. 

The sources that inspire the design of test charters are manifold (cf. \cite{bach2007rapid}\cite{hendrickson2014explore}\cite{kaner2008lessons}), such as risks, product analysis, requirements, and questions raised by stakeholders. Mission statements, test priorities, risk areas, test logistics, and how to test are example elements of a test charter design identified from the literature review and their description~\cite{afzal2015experiment}~\cite{bach2007rapid}~\cite{ghazi2014testing}. Our study will further complement the contents of test charters as they are used in practice.

\section{Research method}\label{sec:rm}

\emph{Study Purpose and Research Questions:} The goal of this study is to investigate the design of test charters and the factors influencing the design of these charters and their contents.

\textit{RQ1: What are the factors influencing the design of test charters?} The factors provide the contextual information that is important to consider when designing test charters, and complements the research on context aware testing~\cite{bach2007rapid}.

\textit{RQ2: What do practitioners include in their test charters?} The checklist of contents supports practitioners to make informed decisions about which contents to include without overlooking relevant ones.

\emph{Interviews:} Interviews (three face-to-face and six through Skype) were conducted with a total of nine industry practitioners through convenience sampling combined with choosing experienced subjects who are visible in the communities discussing ET (see Table \ref{interviewee data}). 

\begin{table}[!b]
\centering
	\caption{Profile of the Interviewees}
	\scalebox{0.7}
	{
\begin{tabular}{l l l l }
	\toprule
 Interview ID & Role & Experience in testing & Organizational size  \\ 
 \midrule
	1                                      & Senior Systems Test Engineer                      & 4 years                                                                                                     & More than 500                                                                                          \\ 
	2                                      & Test Quality Architect                            & 10 years                                                                                                    & 50-500                                                                                                 \\ 
	3                                      & Test Specialist                                   & 10 years                                                                                                    & 50-500                                                                                                 \\ 
	4                                      & Test Consultant                                   & 12 years                                                                                                    & More than 500                                                                                          \\ 
	5                                      & Test Strategist                                   & 3 years                                                                                                     & Less than 50                                                                                           \\ 
	6                                      & CEO, Test Consultant                              & 30 years                                                                                                    & More than 500                                                                                          \\ 
	7                                      & Test Manager                                      & 20 years                                                                                                   & More than 500                                                                                          \\ 
	8                                      & CEO, Test Lead                                    & 4 years                                                                                                    & 50-500                                                                                                 \\ 
	9                                      & Test Quality Manager                              & 13 years                                                                                                   & 50-500                                                                                                 \\ 
	\bottomrule
\end{tabular}
}
	\label{interviewee data}
\end{table}

The interviews were semi-structured, following the structure outlined below:: 
\begin{enumerate}
\itemsep0em
\item \textit{Introduction to research and researcher:} The researchers provide a brief introduction
about themselves, followed by a brief description on the research objectives.
\item \textit{Collection of general information:} In this stage, the information related
to the interviewee is collected . 
\item \textit{Collection of research related information:} This is the last stage where the factors and contents of test charters have been elicited.
\end{enumerate}


\emph{Data analysis:} All the interviews were recorded by consent of the interviewees and later transcribed manually. The qualitative data collected using literature review and interviews was later analyzed using thematic analysis \cite{christ1975review}. After thoroughly studying the coded data, similar codes have been grouped to converge their meaning to form a single definite code. 

\emph{Validity:} The potential bias introduced by interviewing thought leaders and experienced people in the area who are favorable towards exploratory testing may bias the results, and hence may not be fully generalizable. Though, we have not put any value on the factors and contents elicited, and they may be utilized differently depending on context. That is, identifying the potential elements to include in test charters is the first step needed.  To reduce the threat multiple interviews have been used. Using a systematic approach to data analysis (thematic analysis) also aids in reducing this threat.

\section{Results}
\label{sec:results}

\emph{RQ1: What are the factors influencing the design of test charters?} Based on interviews with test practitioners, 30 different factors have been identified (see Table \ref{rq2 codes description}). The table provides the name of the factors as well as a short description of what the factor means. 

\begin{table}[!t]
\caption{Factors influencing test charter design}  
\centering
\scalebox{0.7}
{
\begin{tabular}{l p{12.3cm}}
	\toprule
	\textbf{Charter influence factors} & \textbf{Description}               \\ 
	\midrule
	\textit{F01:} Client Requirements                             & Requirements elicited from clients.                                                                                                                                   \\ 
		\textit{F02:} Test Strategy                                   & Set of ideas that guide the test plan.                                                                                                                                      \\ 
			\textit{F03:} Knowledge of Previous Bugs                      & Knowledge regarding system related bugs that occurred in the past                                                                                                    \\ 
	\textit{F04:} Risk Areas                                     & Results of product risk analysis.                                                                                                                                               \\ 
	\textit{F05:} Time-frame                                       & Time needed for test mission execution, time constraints                                                                                                    \\ 
	\textit{F06:} Project Purpose                                 & Purpose of the project                                                                                                                                            \\ 
	\textit{F07:} Test Function Complexity                        & Complexity of the tested functions.                                                                                                                               \\ 
		\textit{F08:} Functional Flows                                & Flow of data and functions                                                                                                                            \\ 
		\textit{F09:} Product Purpose                                 & Principle goal(s) of the product                                                                                                                                    \\ 
		\textit{F10:} Business Use-case                                & Business use-case for the system.                                                   \\ 
		\textit{F11:} Test Equipment Availability                     & Accessibility to tools and equipment needed for the software tests.                                                                              \\ 
		\textit{F12:} Effort Estimation                               & Effort needed to carry out the test mission.                                                                                                                         \\ 
		\textit{F13:} Test Planning Checklist                         & Testing heuristics appointed for the particular test charter.                                                                                                         \\ 
		\textit{F14:} Product Characteristics                         & Features of the product.                                                                                                                                         \\ 
			\textit{F15:} Quality Requirements                            &Quality requirements of the product.                                                                                        \\ 
			\textit{F16:} Test Coverage Areas                             & Parts of the system to be tested.                                                                      \\ 
				\textit{F17:} Test Team Communication                         & Means of communication between the testing team members.                                                                                                                 \\ 
				\textit{F18:} Project Plan                                    & Plan for the project  prior to its execution.                                                                                                                           \\ 
				\textit{F19:} General Software Design                         & Design of the system software.  \\ 
				\textit{F20:} System Architecture                                  & Structure, interfaces and platforms of the system.                                                                                 \\ 
				\textit{F21:} Process Maturity Level                           & Maturity of the process (e.g. CMMI levels). \\ 
				\textit{F22:} Product Design Effects                         & Impact of product design and features on other modules.                                                                                                              \\ 
	\textit{F23:}Feedback and Consolidation                      & Feedback and consolidation of the test plan based on the comments of previous testers and clients.                                                                      \\ 
	\textit{F24:} Session Notes                                   & Notes filled during previous test sessions.                                                                                                              \\ 
	\textit{F25:} SDLC Phase                                      & Phase involved in the system development life-cycle                                                                                                                       \\ 
	\textit{F26:} Tester                                          & Testers and their experience level                                                                                                                                 \\ 
	\textit{F27:} Client Location                                 & Location of the client, local or global                                                                                                                              \\ 	
	\textit{F28:} System heterogeneity                        & Differences between interacting systems (different programming languages, platforms, system configuration)                                    \\ 
	\textit{F29:} Project Revenue                                 & Business returns for project                                                                                                                                            \\ 
	\textit{F30:} User Journey Map                                & User interaction with the product over time.                                                                                                                            \\
	\bottomrule
\end{tabular}}
\label{rq2 codes description}
\end{table}

We categorized the factors and identified the following emerging categories, namely:
\begin{itemize}
\itemsep0em
\item \textit{Customer and requirements factors:} These factors characterize the customer and their requirements. They include: F01: Client Requirements,
F10: Business Usecase,
F15: Quality requirements,
F27: Client location, and
F30: User Journey Map.
\item \textit{Process factors:} Process factors characterize the context of the testing in regard to the development process. They include: F21: Process Maturity Level and F25: SDLC Phase.
\item \textit{Product factors:} Product factors describe the attributes of the product under test, they include: "F08: Functional flows,
F09: Product Purpose,
F14: Product Characteristics,
F19: General Software design,
F20: System Architecture,
F22: Product Design Effects, and
F28: Heterogeneous Dimensions. 
\item \textit{Project management factors:} These factors concern the planning and leadership aspects of the project in which the testing takes place. They include: F05: Timeframe,
F06: Project Purpose,
F12: Effort estimation,
F17: Test Team Communication,
F18: Project Plan, and
F29: Project Revenue.
\item \textit{Testing:} Testing factors include contextual information relevant for the planning, design and execution of the tests. They include:  F02: Test Strategy,
F03: Knowledge of Previous Bugs,
F04: Risk Areas,
F07: Test Function Complexity,
F11: Test Equipment Availability,
F13: Test Planning Checklist,
F16: Test coverage areas,
F23: Feedback and Consolidation ,
F24: Session Notes, and
F26: Tester.
\end{itemize}

\emph{RQ2: What do practitioners include in their test charters?} The interviews revealed 35 different contents that may be included in a test charter. Table \ref{rq3 codes description} states the content types and their descriptions. 

\begin{table}[!t]
\caption{Contents of test charters}
\centering
\scalebox{0.7}
{
\begin{tabular}{l p{12.3cm}}
	\toprule
	Content type & Description                                                                                                                   \\ 
	\midrule
		C01: Test Setup                                      & Description of the test environment.                                                                                                                              \\ 
		C02: Test Focus                                      & Part of the system to be tested.                                                                                                  \\ 
		C03: Test Level                                      & Unit, Function, System test, etc.                                                                                                      \\ 
		C04: Test Techniques                                 & Test techniques used to carry out the tests.                                      \\ 
		C05: Risks                                           & Product risk analysis.                                                                                                                         \\ 
		C06: Bugs Found                                      & Bugs found previously.                                                                                                                            \\ 
		C07: Purpose                                         & Motivation why the test is being carried out.                                                                                                               \\
		C08: System  Definition                              & Type of system (e.g. simple/ complex).                                                                                                    \\ 
		C09: Client Requirements                             & Requirements specification of the client.                                                                                                        \\ 
		C10: Exit Criteria                                   & Defines the ``done'' criteria for the test.                                                                                                             \\
		C11: Limitations                                     & It tells of what the product must never do, e.g. data sent as plain text is strictly forbidden.                                                              \\ 
		C12: Test Logs                                       & Test logs to record the session results.                                                                                                                            \\ 
		C13: Data and Functional Flows                       & Data and work flow among components.                                                                                    \\ 
		C14: Specific Areas of  Interest                     & Where to put extra focus on during the testing.                                               \\ 
		C15: Issues                                          & Charter specific issues or concerns to be investigated.                                                    \\ 
		C16: Compatibility Issues                            & Hardware and software compatibility and interoperability issues.                                                                                      \\ 
		C17: Current Open Questions                          & Existing questions that refer to the known unknowns.                                                                                                   \\ 
		C18: Information Sources                             & Documents and guidelines that hold information regarding the features, functions and systems being tested.                                 \\
		C19: Priorities                                      & Determines what the tester spends most and least time on.                                           \\ 
		C20: Quality Characteristics                         & Quality objectives for the project.                                               \\ 
		C21: Test Results Location                           & Test results location for developers to verify.                                                                                                   \\ 
		C22: Mission Statement                               & One liner describing the mission of the test charter.                                                                                           \\ 
		C23: Existing Tools                                  & Existing software testing tools that would aid the tests.                                                                                     \\ 
		C24: Target                                          & What is to be achieved by each test.                                                                                                                       \\ 
		C25: Reporting                                       & Test session notes.                                                                                                                                 \\ 
		C26: Models and Visualizations                       & People, mind maps, pictures related to the function to be tested.                                                                                 \\ 
		C27:  General Fault                                   & Test related failure patterns of the past.                                                                                                                 \\ 
		C28: Coverage                                        & Charter’s boundary in relation to what it is supposed to cover.                                                                                   \\ 
		C29: Engineering Standards                           & Regulations, rules and standards used, if any.                                                                                                         \\ 
		C30: Oracles                                         & Expected behavior of the system (either based on requirements or a person) \\ 
		C31: Logistics                                       & How and when resources are used to execute the test strategy, e.g. how people in projects are coordinated and assigned to testing tasks.  \\ 	 
		C32: Stakeholders                                    & Stakeholders of the project and how their conflicting interests would be handled.                                                        \\ 
		C33: Omitted Things                                  & Specifies what will not be tested.                                                                                                                   \\ 
		C34: Difficulties                                    & The biggest challenges for the test project.                                                                                                          \\ 
		C35: System Architecture                             & Structure, interfaces and platforms concerning the system, and its impact on system integration.                                                                                        \\ 
	\bottomrule
	
\end{tabular}}
\label{rq3 codes description}
\end{table}

Similar to the factors we categorized the contents as well. Seven categories have been identified, namely testing scope, testing goals, test management, infrastructure, historical information, product-related information, and constraints, risks and issues. 

\begin{itemize}
\item \textit{Testing scope:} The testing scope describes what to focus the testing on, be it the parts of the system or the level of the testing. It may also describe what not to focus on and set the priorities. It includes: C02: Test Focus,
C03: Test Level, 
C04: Test Techniques,
C10: Exit Criteria,
C14: Specific Areas of  Interest,
C19: Priorities,
C28: Coverage, and
C33: Omitted Things.
\item \textit{Testing goals:} The testing goals set the mission and purpose of the test session. They include: C07: Purpose,
C22: Mission Statement, and
C24: Target.  
\item \textit{Test management:} Test management is concerned with the planning, resource management, and the definition of how to record the tests. Test management includes: C12: Test Logs, 
C18: Information Sources, 
C21: Test Results Location, 
C25: Reporting,
C26: Models and Visualizations, 
C31: Logistics,
C32: Stakeholders, and
C34: Difficulties.
\item \textit{Infrastructure:} Infrastructure comprises of tools and setups needed to conduct the testing. It includes: C01: Test Setup and
C23: Existing Tools.
\item \textit{Historical information:} As exploratory testing focuses on learning, past information may be of importance. Thus, the historical information includes: C06: Bugs Found,
C16: Compatibility Issues, 
C17: Current Open Questions, and
C27: General Fault.
\item \textit{Product-related information:} Here contextual product information is captured, including: C08: System Definition,
C13: Data and Functional Flows, and
C35: System Architecture. 
\item \textit{Constraints, risks and issues:} Constraints, risks and issues to testing comprise of the items: C05: Risks,
C15: Issues, and
C29: Engineering Standards.
\end{itemize}

\section{Conclusion}\label{sec:conclusion}

In this study two checklists for test charter design were developed. The checklists were based on nine interviews. The interviews were utilized to gather a checklist for factors influencing test charter design and one to describe the possible contents of test charters. Overall, 30 factors and 35 content types have been identified and categorized.

The factors may be used in a similar manner and should be used to question the design choices of the test charter. For example: 
\begin{itemize}
\itemsep0em
\item Should the test focus of the charter be influenced by previous bugs (F03)? How/why? 
\item Are the product's goals (F09) reflected in the charter? 
\item Is it possible to achieve the test charter mission in the given time for the test session (F12)?
\item etc.
\end{itemize}

With regard to the content a wide range of possible contents to be included have been presented. For example, only stating the testing goals (C22) provides much room for exploration, while adding the techniques to be used (C04) may constrain the tester. Thus, the more information is included in the test charter the exploration space is reduced. Thus, when deciding what to include from the checklist (Table \ref{rq3 codes description}) the possibility to explore should be taken into consideration. 

In future work we need to empirically understand (a) which are the most influential factors and how they affect the test charter design, and (b) which of the identified contents should be included to make exploratory testing effective and efficient.

\bibliographystyle{abbrv}

\end{document}